\documentclass[11pt,a4paper]{article}
\usepackage{jheppub}
\usepackage{cases}
\usepackage{graphicx}
\usepackage{dcolumn,color}
\usepackage{bm}
\usepackage{epstopdf}
\usepackage{epsfig}
\usepackage{color,framed}
\usepackage{amsmath,amsthm,amssymb} 

\usepackage[table]{xcolor}

\title{Brane structure and metastable graviton in five-dimensional model with (non)canonical scalar field}

\author[a,b]{Yuan Zhong,}
\author[a,1]{Yu-Xiao Liu,\note{The corresponding author.}}
\author[c]{Zhen-Hua Zhao}

\affiliation[a]{Institute of Theoretical Physics, Lanzhou University,\\
 Lanzhou 730000, People' s Republic of China}
\affiliation[b]{IFAE, Universitat Aut\`onoma de Barcelona, 08193 Bellaterra, Barcelona, Spain}
\affiliation[c]{Department of Applied Physics, Shandong University of Science and Technology,\\ Qingdao 266590, People' s Republic of China}

\emailAdd{yzhong@ifae.es, liuyx@lzu.edu.cn, zhaozhh09@lzu.edu.cn}

\abstract{The appearance of inner brane structure is an interesting issue in domain wall {brane model}. Because such structure usually leads to quasilocalized modes of various kinds of bulk fields. In this paper, we construct a domain wall brane model by using a scalar field $\phi$, which couples to its kinetic term. The inner brane structure emerges as the scalar-kinetic coupling increases. With such brane structure, we show that it is possible to obtain gravity resonant modes in both tensor and scalar sectors. The number of the resonant modes depends on the vacuum expectation value of $\phi$ and the form of scalar-kinetic coupling. The correspondence between our model and the canonical one is also discussed. The noncanonical and canonical background scalar fields are connected by an integral equation, while the warp factor remains the same. Via this correspondence, the canonical and  noncanonical models share the same linear perturbation spectrum. So the gravity resonances {obtained} in the noncanonical frame can also be obtained in the standard model. However, due to the inequivalence between the corresponding background scalar solutions, the localization condition for the left-chiral fermion zero mode can be largely different in different frames. Our estimate shows that the magnitude of the Yukawa coupling in the noncanonical frame might be hundreds times larger than the one in the canonical frame, if one demands the localization of the left-chiral fermion zero mode as well as the appearance of a few gravity resonance modes.}

\keywords{Field Theories in Higher Dimensions, Large Extra
Dimensions}

\begin{document}
\maketitle


\section{Introduction}
The idea of the existence of extra dimensions was proposed even before Einstein's general relativity. In 1914, the Finnish physicist Gunnar Nordst\"om applied an extra spatial dimension to unify his own theory of gravity and Maxwell's electromagnetic theory. The most well known extra dimension theory was proposed by Kaluza and then developed by Klein in the 1920s. Kaluza-Klein {(KK)} theory unified general relativity with Maxwell's theory by introducing a compact extra dimension. With the development of Yang-Mills theory, physicists tried to introduce more extra dimensions to unify non-Abelian gauge fields with general relativity. More ambitiously, they tried to use extra dimensions to unify matter fields and gauge fields in a single theory, for example, the superstring theory or higher dimensional supergravity theories (see~\cite{AppelquistChodosFreund1987} for the history and early development of extra dimension theories). One of the common feature of the early proposals for extra dimensions is that the extra dimensions are compacted to scales that are too tiny to detect. Besides, unification is always the main motivation for introducing extra dimensions.

This situation begins to change in 1980s, when physicists realized that large extra dimensions are possible if matter fields are trapped on four-dimensional sub-manifolds~\cite{Akama1982,RubakovShaposhnikov1983,Visser1985}. Large extra dimensions are a general prediction of some perturbative string theories~\cite{Antoniadis1990,AntoniadisArkani-HamedDimopoulosDvali1998}. Later on, it was found that {extra dimensions} can be an alternative solution to the hierarchy problem in the standard model of particle physics~\cite{Arkani-HamedDimopoulosDvali1998a,RandallSundrum1999}. Another astonishing discovery in 1999 is that even if we are living with an infinitely large extra dimension, we would still observe effectively a four-dimensional Newtonian gravity, provided the space-time is nonfactorizable and is properly warped~\cite{RandallSundrum1999a}. The model in ref.~\cite{RandallSundrum1999a} is now referred to as the Randall-Sundrum-2 (RS2) model, which assumes that we  live on a 3-brane embedded in an AdS$_5$ space. {The spectrum of KK gravitons} of the RS2 model is constituted by a normalizable zero mode and a continuum of gapless massive modes. As usual, the zero mode is responsible for the reproduction of four-dimensional Newtonian gravity. However, in contrast to the traditional models with factorizable geometry, now the massive modes only cause a small correction to Newtonian gravity, even there are extremely light modes. The reason is that the couplings of these massive modes to matter on the brane are sufficiently suppressed, so that the integration over all of them only gives a subleading contribution to the gravitational interaction between two test masses on the brane.

It is interesting to search for a nonsingular thick version of the RS2 model, especially, to see if gravity can also be localized on a nonsingular domain wall~\cite{KehagiasTamvakis2001,Gremm2000}, rather than on {an infinitely} thin brane. The domain wall brane in ref.~\cite{KehagiasTamvakis2001} is an extension of ref.~\cite{RubakovShaposhnikov1983} in warped space-time. While the solution in ref.~\cite{Gremm2000} takes the advantage in analytical computations. One of the important motivations for studying thick versions of the RS2 model is that the graviton spectrum might be nontrivially different from the original RS2 model. For example, scalar modes begin to contribute to gravitational interactions~\cite{Giovannini2001a,Giovannini2003}. Besides, in thick brane models it is possible to find graviton resonances in both tensor and scalar sectors.

The possibility of existing graviton resonances in thick brane model was first noted by Gremm~\cite{Gremm2000}. In the model of ref.~\cite{Gremm2000}, gravitons are trapped by a volcanolike potential which asymptotically approaches to zero at the infinity of extra dimension. So, the only bound state is the zero mode. However, the special shape of the graviton potential indicates the possibility for finding massive resonant modes. {These modes} would give a quasidiscrete spectrum of low mass KK {modes} with unsuppressed couplings to matter on the brane. So, the existence of such resonant modes would probably change the four-dimensional physics. For example, when the tensor zero mode is quasilocalized rather than localized, the effective four dimensions are preserved only in an intermediate scale, while in both ultra large and ultra {small scales} the gravity is five-dimensional~\cite{DvaliGabadadzePorrati2000a,GregoryRubakovSibiryakov2000a,Cs'akiErlichHollowood2000,ShaposhnikovTinyakovZuleta2004,CveticRobnik2008}. Unfortunately, the solution of ref.~\cite{Gremm2000} does not support any graviton resonance. Studies on the linear structure of other thick brane~\cite{KehagiasTamvakis2001} also concluded the absence of narrow graviton resonance (see~\cite{ShaposhnikovTinyakovZuleta2005,BazeiaFurtadoGomes2004}).

Recently, it was noticed that the inner brane structure {plays} a crucial role in generating graviton resonant modes~\cite{GuoLiuZhaoChen2012,XieYangZhao2013,CruzSousaMalufAlmeida2014}.
However, the model of refs.~\cite{XieYangZhao2013,CruzSousaMalufAlmeida2014} contains two background scalar fields, which may make the model potentially problematic. Because in a model with multi scalar fields, there might be a normalizable scalar zero mode that will transmit a new force we have never seen before, so is phenomenologically unacceptable~\cite{George2011}. As to the model of ref.~\cite{GuoLiuZhaoChen2012}, the nonminimal coupling between the background scalar and gravity makes it is very hard to obtain analytical solutions, needless to say the whole spectrum of linear perturbations. So far, the stability of the solution in ref.~\cite{GuoLiuZhaoChen2012} against scalar perturbations is still unclear.

It is interesting to search for a single scalar field thick brane model, which supports inner brane structure and graviton resonances.
For one thing, the linearization of a large class of single scalar field model has been studied in ref.~\cite{ZhongLiu2013}, where the matter Lagrangian density is $\mathcal{L}(\phi,X)$ with $X=-\frac12\partial_M\phi\partial^M\phi$ {the} kinetic term of the background scalar field $\phi$. With such a matter Lagrangian density, the scalar $\phi$ can have noncanonical kinetic terms. For this reason, $\phi$ is also dubbed as the $K$-field. The $K$-field was initially introduced in cosmology as a new mechanism of inflation ~\cite{Armendariz-PiconDamourMukhanov1999,GarrigaMukhanov1999,Armendariz-PiconMukhanovSteinhardt2001},
and later was applied in brane models ~\cite{BazeiaBritoNascimento2003,KoleyKar2005b,AdamGrandiKlimasSanchez-GuillenWereszczynski2008,BazeiaGomesLosanoMenezes2009,LiuZhongYang2010,CastroMeza2013}.
The linearization of a brane model not only allows us to study the stability of the solution, but also helps us to analyze the structure of the graviton spectrum.

Besides, by using the superpotential method, one can obtain some interesting analytical solutions for $K$-field models. For example, the case with $\mathcal{L}(\phi,X)=X+\alpha X^2$ has been studied first in ref.~\cite{BazeiaGomesLosanoMenezes2009} for small $\alpha$, and then in ref.~\cite{ZhongLiuZhao2013} for arbitrary positive $\alpha>0$. In ref.~\cite{ZhongLiuZhao2013}, we used a different superpotential method, so that analytical thick brane solution can be easily obtained even for very large $\alpha$. However, as shown in~\cite{ZhongLiuZhao2013}, there is no sign for any graviton resonance either in the tensor or {scalar} sector. So it is worth to try other types of noncanonical terms.

In this paper, we investigate {the model of a scalar field} which couples to its own kinetic term, so the matter Lagrangian density {takes} the form: $\mathcal{L}=G(\phi)X - V(\phi )$, where $G(\phi)$ and $V(\phi)$ are arbitrary functions of $\phi$. This model will be set up and solved in the next section. We will show how inner brane {structure emerges} with the noncanonical kinetic term. Then, in section~\ref{sec3}, we discuss the stability of our solution against tensor and scalar perturbations. In section~\ref{sec4}, we use numerical method to solve the linear perturbation equations, and find out the possible graviton resonances in both tensor and scalar {sectors}. In section~\ref{sec5}, we show how to obtain all the good properties of the noncanonical brane solution in a canonical model. The relation between the {canonical and noncanonical models}  will also be addressed. In section~\ref{sec6}, we consider the {localization} of massless fermion in both the canonical and noncanonical frames, and discuss how the noncanonical term affects the localization.

\section{The emergence of brane structure}
\label{sec2}
We study a model with the following action:
\begin{eqnarray}
S=\int d^5 x \sqrt{-g}\left(\frac{R}{2\kappa_5^2}
+G(\phi)X  - V(\phi )\right),
\end{eqnarray}
where $g=\det (g_{MN})$ ($M, N=0,1,2,3,5$ are indices of the bulk coordinates) and $\kappa_5^2$ denotes the five-dimensional gravitational constant.
The metric is taken to be
\begin{eqnarray}
\label{metric}
ds^2 = \textrm{e}^{2A(y)}\eta_{\mu\nu}dx^\mu dx^\nu + dy^2,
\end{eqnarray}
with $\eta_{\mu\nu}=\textrm{diag}(-1,+1,+1,+1)$ the four-dimensional Minkowski metric, and $\textrm{e}^{2A(y)}$ the warp factor. Here, $y\equiv x^5$ denotes the extra dimension,
$\mu, \nu, \cdots$ are indices of {the} brane coordinates.

The independent dynamical equations for the system are the Einstein equations:
\begin{subequations}
\label{Eqy}
\begin{eqnarray}
\label{eqy1}
-3\partial_y^2 A& =& \kappa _5^2{\mathcal {L}_X}(\partial_y \phi)^2,\\
\label{eqy2}
6(\partial_y A)^2& =& \kappa _5^2({\cal L}+{\mathcal {L}_X}(\partial_y \phi)^2).
\end{eqnarray}
\end{subequations}
We use subscripts of $\mathcal {L}$ to represent derivatives of $\mathcal {L}$ with respect to corresponding arguments, for example, $\mathcal {L}_X= \partial \mathcal {L}/\partial X$. From now on, let us focus on a model with $G(\phi)=1+\beta\phi^{2n}$ and $n=1,2,\cdots$. Since $\beta=0$ corresponds to the standard model of a thick brane~\cite{KehagiasTamvakis2001,Gremm2000}, let us call $\beta$ the deviation parameter.

To find the exact and analytical solution of the Einstein equations \eqref{Eqy}, we follow the procedures proposed in~\cite{ZhongLiuZhao2013} and assume
\begin{eqnarray}
\label{AW}
\partial_y A&=&-\frac{\kappa_5^2}{3}(W(\phi)+\beta Z(\phi)),\\
\label{phiW}
\partial_y \phi&=&{W_\phi },
\end{eqnarray}
where $W$ and $Z$ (called the superpotentials) are functions of $\phi$, and $W_\phi\equiv\frac{dW}{d\phi}$.
Plugging the above equations into eq.~\eqref{eqy1} and comparing the coefficients of $\beta$, we immediately obtain
\begin{eqnarray}
\label{Zphi}
Z_\phi = {\phi^{2n}}W_\phi.
\end{eqnarray}
From {eq.}~\eqref{eqy2}, we get
\begin{eqnarray}
\label{VW}
V = \frac{1}{2}{W_\phi }^2  + \frac{1}{2}\beta {\phi ^{2n}}{W_\phi }^2 - \frac{2}{3}\kappa _5^2{(W + \beta Z)^2}.
\end{eqnarray}
Therefore, given a $W(\phi)$, an analytical solution can be obtained by solving two first-order differential equations \eqref{AW} and \eqref{phiW} {with the} constraint equation~\eqref{Zphi}.

The superpotential method allows us to find some analytical solutions.
For simplicity, let us consider a cubic superpotential
\begin{eqnarray}
W=k \phi_0^2\left(\frac{\phi }{\phi_0}-\frac{1}{3}\left(\frac{\phi }{\phi_0}\right)^3\right).
\end{eqnarray}
Inserting $W$ into eq.~\eqref{phiW}, we obtain the standard kink configuration for the background scalar field:
\begin{eqnarray}
\label{phi}
\phi=\phi_0\tanh(k y),
\end{eqnarray}
and from eq.~\eqref{Zphi}, we immediately get
\begin{eqnarray}
Z =-\frac{k}{(3+2 n) \phi_0}\phi ^{3+2 n}+\frac{k \phi_0}{1+2 n}\phi ^{1+2 n}.
\end{eqnarray}
Finally, {the} scalar potential is given by
\begin{eqnarray*}
V&=&\frac{k^2\phi _0^2}{6} \left[3 \left(\frac{\phi ^2}{\phi _0^2}-1\right)^2\left(1+\beta  \phi ^{2 n}\right)\right.\nonumber\\
&-&\left.
\frac{4}{9} \kappa _5^2 \phi_0^2  \left(\frac{\phi ^2}{\phi _0^2}+\frac{3 \beta  \phi ^{2 n}}{3+2 n}\frac{\phi ^2}{\phi _0^2}-3-\frac{3 \beta  \phi ^{2 n} }{1+2 n}\right)^2\right].
\end{eqnarray*}
By taking 
$A(0)=0$, we obtain the general solution for the warp factor:
\begin{eqnarray}
\label{warpfactorSolution}
A&=&\frac{1}{18} \kappa _5^2 \phi _0^2\left[ -\tanh^2(ky) -4\ln (\cosh (k y))\right.
\nonumber\\&-&
\left.\frac{{\beta {c_n}\big(1 + 2n + 2{\cal H}(n,y)\big)}}{{(1 + n)}}{\tanh ^{2(n + 1)}}(ky)\right],
\end{eqnarray}
where
$c_n$ is defined by
\begin{eqnarray}
c_n=\frac{3  \phi_0^{2 n} }{(1+2 n) (3+2 n)},
\end{eqnarray}
and the function $\mathcal{H}(n,y)$ is
\begin{eqnarray}
\mathcal{H}(n,y)=F_2^1\left[1,1+n;2+n;\tanh ^2(k y)\right],
\end{eqnarray}
with $F_2^1\left[a,b;c;w\right]$ {the hypergeometric function}.
When $\beta=0$, our solution reduces to the one given in~\cite{KehagiasTamvakis2001}.
For simplicity, from now on, let us take the dimensionless quantity $\kappa _5^2 \phi _0^2=\frac 32$, and consider $\phi_0$ as a parameter.

\begin{figure*}
\begin{center}
\includegraphics[width=0.45\textwidth]{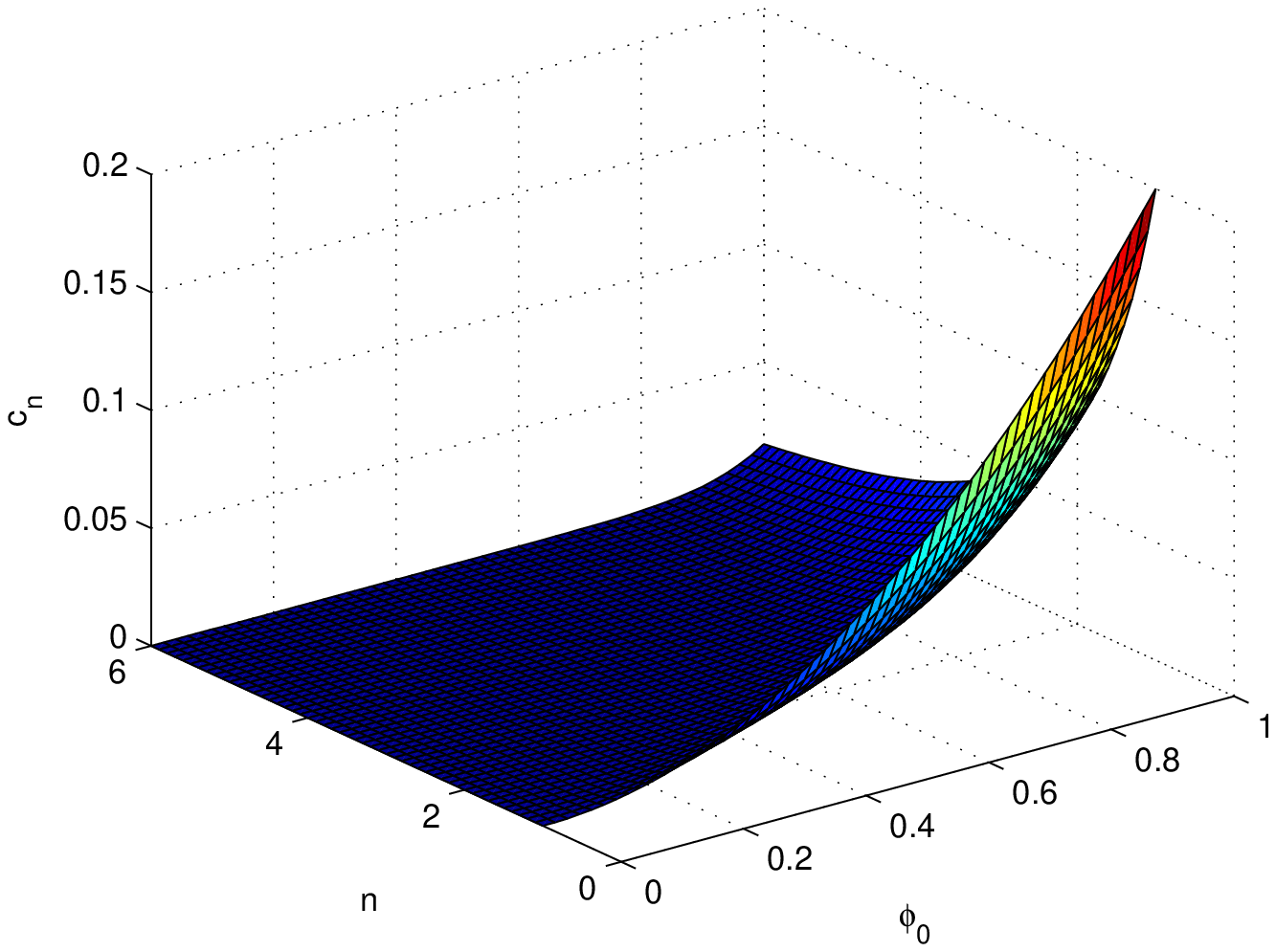}
\includegraphics[width=0.45\textwidth]{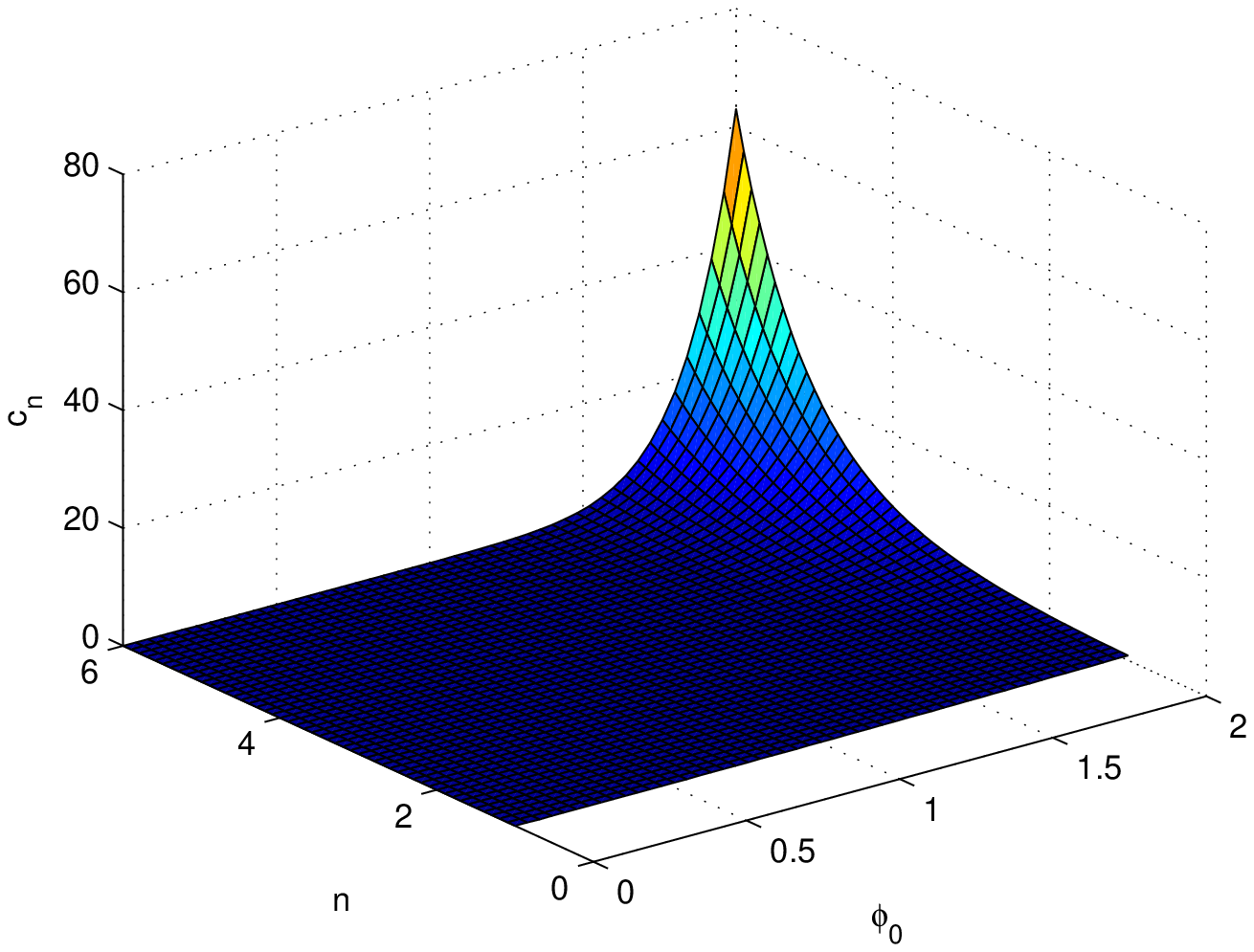}
\end{center}
\caption{The behaviour of the coefficient $c_n$ with parameters $\phi_0>2$ {and $n$}. When $\phi_0\in(0,1]$, $c_n$ decreases as $n$ increases. While when $\phi_0\geq2$, $c_n$ { increases} rapidly as $n$ increases.}
\label{figureCoefficient}
\end{figure*}

The asymptotic behavior of $A$ at $|y|\to \infty$ is
\begin{equation}
\label{Aasyp}
A \to -\left(1+ \beta c_n\right)k |y|,
\end{equation}
so, the space-time is asymptotically {anti-de} Sitter. Note that the behavior of $c_n$ depends on the vacuum expectation value of $\phi$, i.e., $\phi_0$. Roughly speaking, when $\phi_0\in(0,1]$, $c_n$ decreases as $n$ increases, while when $\phi_0\geq2$, $c_n$ { increases} rapidly as $n$ increases (see figure~\ref{figureCoefficient}). Since $c_n$ embodies the deviation from RS2 model in large scale, it should also reflects the impact of noncanonical term on the inner brane structure, and finally on the graviton spectrum. So from now on, we assume $\phi_0\geq2$, so that the noncanonical effects become dominant as $n$ increases.

\begin{figure}
\begin{center}
\includegraphics[width=0.6\textwidth]{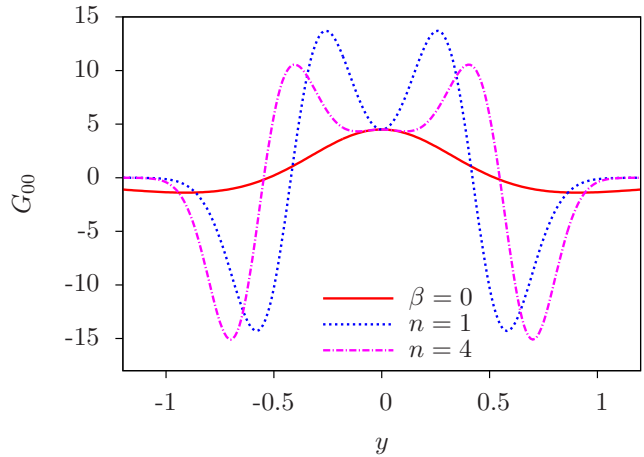}
\end{center}
\caption{Plots of $G_{00}$ for $\beta=0$ (the solid line) and $n=1$ (the dotted line), and $n=4$ (the dotted-dashed line). The parameters are $k=1$, $\phi_0=2$, and for $n=1,4$ we take $\beta=8$.} \label{figureEnergyDensityBeta}
\end{figure}

To see how $\beta$ and $n$ affect the inner brane {structure}, let us consider the zero-zero component of the Einstein tensor $G_{MN}=R_{MN}-\frac12g_{MN}R$:
\begin{eqnarray}
G_{00}=-3 \textrm{e}^{2 A}\left[2 \left(\partial _yA\right)^2+\partial _y^2 A\right].
\end{eqnarray}
From figure~\ref{figureEnergyDensityBeta}, we see that when $\beta=0$, $G_{00}$ has only one peak {around $y=0$}. However, when $\beta\neq0$, the original peak begins to deform into two (for $n=1$) or three (for $n\geq 2$) peaks, which implies the emergence of brane {structure} in the vicinity of $y=0$.

Branes with inner {structure} are theoretically interesting, because they can be regards as the smooth versions of the Lykken-Randall model~\cite{LykkenRandall2000}. Although the sub-branes in our model is symmetric, one can generate asymmetric sub-branes by choosing appropriate superpotential $W(\phi)$. It is interesting to note that in some models with asymmetric sub-branes, fermion is localized on one of the sub-branes, while gravity on another~\cite{AhmedGrzadkowski2012,XieYangZhao2013}. Hopefully, the hierarchy problem can be solved in such models (see~\cite{AhmedGrzadkowski2012} for related discussions).

Inner brane {structure was} originally constructed in models with two real scalar fields~\cite{BazeiaGomes2004}, or a complex scalar field~\cite{Campos2002,ZhaoLiuWangLi2011}. In these models, the energy density of brane deforms as some parameters vary, as a result branes {acquire inner structure}. However, as stated in ref.~\cite{George2011}, localizable scalar zero modes might survive in models with more than one real scalar field, and therefore, is phenomenologically unacceptable.

Inner brane {structure} can also be realized either in general relativity~\cite{BazeiaFurtadoGomes2004}, or in modified theories of gravity~\cite{GuoLiuZhaoChen2012,YangLiZhongLi2012,LiuChenHeng-GuoZhou2012,BazeiaLobaoMenezesPetrovSilva2013} by using only a single scalar field. But the linearization of gravity beyond general relativity is rather evolve, especially, when scalar perturbations are considered.
On the contrary, the model we constructed above not only circumvents the {problem confronted by} models with multi scalar fields, but also has a simple linear structure, which allows us to analyze the graviton resonances.

In what follows, it is more convenient to work in the {conformally} flat coordinates:
\begin{eqnarray}
\label{metricConformal}
ds^2 = \textrm{e}^{2A(r)}(\eta_{\mu\nu}dx^\mu dx^\nu + dr^2),
\end{eqnarray}
where $dr\equiv \textrm{e}^{-A} dy$. The {derivative} with respect to $r$ will be denoted by a prime, for example, $A'\equiv\partial_r A$.

\section{Linear stability}
\label{sec3}
In this section, we examine the stability of the aforementioned brane solution under small field perturbations $\{\delta g_{MN},~\delta\phi\}$. In $r$-coordinate, it is much convenient to define the metric perturbations as $\delta g_{MN}\equiv \textrm{e}^{2A(r)}h_{MN}$. Both $h_{MN}$ and $\delta\phi$ are functions of the bulk coordinates $(x^\rho, r)$. In order to derive the master equations for the linear perturbations, we usually introduce the scalar-tensor-vector decomposition. After this decomposition, the original perturbations can be classified in to scalar, tensor and vector modes. Each type of modes evolves independently. The linear perturbation equations for a general class of $K$-brane models have been derived in~\cite{ZhongLiu2013}. According to~\cite{ZhongLiu2013}, the spectrum of the vector modes contains only a nonlocalizable zero mode. So, we omit the vector modes and only give a brief review on the tensor and scalar modes here.

\subsection{Tensor mode}
For the tensor mode, the perturbed metric is
 \begin{eqnarray}
ds^2 = \textrm{e}^{2A(r)}\left[(\eta_{\mu\nu}+D_{\mu\nu})dx^\mu dx^\nu+dr^2\right],
\end{eqnarray}
where $D_{\mu\nu}$ is the transverse and traceless tensor perturbation, which satisfies the following equation~\cite{ZhongLiu2013}:
\begin{eqnarray}
  \label{tens}
\square ^{(4)}D_{\mu \nu }+ D_{\mu \nu }'' + 3A'D_{\mu \nu }'&=& 0,\quad
\square ^{(4)}=\eta^{\mu\nu}\partial_\mu\partial_\nu.
\end{eqnarray}
This is the same equation that $D_{\mu\nu}$ obeys in the standard model~\cite{DeWolfeFreedmanGubserKarch2000}. In other words, the {introduction}  of noncanonical kinetic terms does not affect the structure of the tensor perturbation equation. To understand this, let us note that the tensor mode is decoupled from the scalar modes, and in our model we only modify the scalar Lagrangian of the standard model. So the dynamical equation for the tensor mode takes the same form as the one in the standard model. On the other hand, in some models that modified the gravity part (for example, in $f(R)$ gravity), the tensor perturbation equation can have nontrivial modifications~\cite{ZhongLiuYang2011,LiuZhongZhaoLi2011}.

To continue, let us introduce the following decomposition:
\begin{eqnarray}
D_{\mu \nu }(x^\rho,r)=\textrm{e}^{-3/2A}\epsilon_{\mu\nu}( x^\rho)\chi(r),
\end{eqnarray}
where $\epsilon_{\mu\nu}( x^\rho)$ is transverse and traceless $\eta^{\mu\nu}\epsilon_{\mu\nu}=0=\partial^\mu\epsilon_{\mu\nu}$ and satisfies $\square ^{(4)}\epsilon_{\mu\nu}=m^2 \epsilon_{\mu\nu}$. Then, the KK mode $\chi(r)$ satisfies a Schr\"odinger-like equation
\begin{eqnarray}
 - \chi'' + {U_T}(r)\chi = {m^2}\chi,
\end{eqnarray}
with
\begin{eqnarray}
\label{SchTensor}
{U_T}(r) = \frac{9}{4}A'^2 + \frac{3}{2}A''.
\end{eqnarray}
This equation can be factorized as
\begin{eqnarray}
\label{tensorFactor}
\mathcal{J}\mathcal{J}^\dagger\chi={m^2}\chi,
\end{eqnarray}
with
\begin{eqnarray}
\mathcal{J}\equiv \partial _r+\frac32A',\quad
\mathcal{J}^\dagger=-\partial _r+\frac32A'.
\end{eqnarray}
According to the supersymmeric quantum mechanics, when the Schr\"odinger-like equation can be factorized as eq.~\eqref{SchTensor}, the corresponding {eigenvalue} must be positive semi-definite, namely, $m^2\geq 0$. The absence of tachyons implies that our solution is stable against tensor perturbation.

\subsection{Scalar modes}
Now, let us verify the stability of our solution against linear scalar perturbations.
It is more convenient to analyze the scalar modes in the longitude gauge. In this gauge the perturbed metric takes the following form:
 \begin{eqnarray}
ds^2 = \textrm{e}^{2A(r)}\left[\eta_{\mu\nu}(1+\Psi)dx^\mu dx^\nu+(1+\Xi)dr^2\right].
\end{eqnarray}
Then the perturbation equations are~\cite{ZhongLiu2013}
\begin{eqnarray}
\label{18}
  && \Xi=- 2\Psi,  \\&&
\label{19}
  \frac{3}{2}A'{\Xi} - \frac{3}{2}\Psi' = \kappa _5^2{\mathcal{L}_{X}}\phi '\Phi,\\
\label{20}
&&\frac{3}{2}{\square ^{(4)}}\Psi - \frac{3}{2}\Psi'' - \frac{3}{2}A'\Psi' = 2\kappa _5^2{\mathcal{L}_{X}}\phi '\Phi ' + \kappa _5^2\phi {'^2}{\mathcal{L}_{X\phi }}\Phi.
\end{eqnarray}
Here $\Phi=\delta\phi$, and we have used the fact that $\mathcal{L}_{XX}=0$. Using eqs.~\eqref{18}, \eqref{19} and the background equations \eqref{Eqy}, one can eliminate $\Xi$, $\Phi$, and $\mathcal{L}_{X\phi}$ in eq.~\eqref{20} and obtain the following equation:
\begin{eqnarray}
 && {\square ^{(4)}}\Psi +\Psi''
  +\left[\partial _r\ln  \left(\frac{\textrm{e}^{3A(r)}}{\mathcal{L}_X(\phi ')^2}\right)\right]\Psi '
  +2 A' \left[\partial _r\ln  \left(\frac{A'^2}{\mathcal{L}_X(\phi ')^2}\right)\right]\Psi=0.
\end{eqnarray}
The above equation takes a more compact form
\begin{eqnarray}
\label{schro}
{\square ^{(4)}}\hat{\Psi}+\hat{\Psi}''- \zeta\left(\zeta^{-1}\right)''\hat{\Psi} =0,
\end{eqnarray}
if one defines
\begin{eqnarray}
\label{30}
\Psi =\textrm{e}^{-3A/2}\mathcal{L}_X^{1/2}\phi '\hat{\Psi},
\end{eqnarray}
and
\begin{eqnarray}
\label{31}
\zeta= \textrm{e}^{3A/2}\frac{\phi '}{A' }\mathcal{L}_X^{1/2}.
\end{eqnarray}
Inserting the KK decomposition
\begin{equation}
\hat{\Psi}=\sum_j \textrm{e}^{ip_\mu^{j} x^\mu}\varphi_{j}(r)
\end{equation}
into eq.~\eqref{schro} and using $(p_\mu^{j})^2=-m_j^2$, one immediately obtain the following equation for $\varphi_{j}(r)$:
\begin{equation}
\label{Psi}
\mathcal{A}^\dagger\mathcal{A} \varphi_{j}(r)=m_j^2\varphi_{j}(r),
\end{equation}
with
\begin{equation}
\label{Adagger}
\mathcal{A}=\frac{d}{dr}+\frac{\zeta'}{\zeta},\quad
\mathcal{A}^\dagger=-\frac{d}{dr}+\frac{\zeta'}{\zeta}.
\end{equation}
Similar to the case of tensor sector, we obtain a factorizable Schr\"odinger-like equation for scalar modes. Therefore, the stability of the solution against scalar perturbations is guaranteed.

Note that the above procedures are valid only if $\mathcal{L}_X>0$. Considering the kink configuration of $\phi$: $\phi(\pm\infty)=\pm\phi_0$, and $\phi(0)=0$, the stability condition reduces to
\begin{eqnarray}
\label{stab}
\beta>-\frac1{\phi_0^{2n}}\equiv\beta_c.
\end{eqnarray}
For simplicity, in our following discussion, we assume $\beta>0$, so that the stability condition is always satisfied.

\section{Mass spectrum of gravitons}
\label{sec4}
In this section, we analyze how the noncanonical kinetic term affects the mass spectrum of gravitons, especially, the localization of zero modes and the appearance of massive gravity resonances. We study the appearance of gravity resonances in both the tensor and scalar sectors, and illustrate how the value of $n$ affects the number of resonances.
\subsection{Zero modes}
The zero modes {correspond} to infinitely long range forces. To reproduce the four-dimensional Newtonian gravity, we require the localization of the {tensor zero mode}. The zero mode of tensor perturbation can be easily read out from eq.~\eqref{tensorFactor}:
\begin{eqnarray}
\chi_0\propto \textrm{e}^{3/2A}.
\end{eqnarray}
The normalization condition for the zero mode is
\begin{eqnarray}
\int dr  \textrm{e}^{3A(r)}=\int dy \textrm{e}^{2A(y)}<\infty.
\end{eqnarray}
This condition is satisfied, if
\begin{eqnarray}
\beta  > - \frac{1}{c_n}=-\frac{(1+2 n) (3+2 n)}{3  \phi_0^{2 n} }\equiv \beta_t.
\end{eqnarray}
Obviously, $\beta_c>\beta_t$, for $n\geq 1$. Therefore, for any solution that satisfies the stable condition $\beta>\beta_c$, the localization condition for {the} tensor zero mode is also satisfied. As a result, four-dimensional Newtonian gravity can be reproduced, provided the stability condition \eqref{stab} is satisfied.

Similarly, the expression for the scalar zero mode can be read out from eq.~\eqref{Psi}:
\begin{eqnarray}
\varphi_0\propto \zeta^{-1}.
\end{eqnarray}
The localization condition is $\int dr \varphi_0^2<\infty$, namely,
\begin{eqnarray}
&&\int dr \textrm{e}^{-3A}\frac{A'^2}{\phi'^2\mathcal{L}_X}
=\int dy \textrm{e}^{-4A}\frac{(\partial_y A)^2}{\mathcal{L}_X(\partial_y\phi)^2}
\nonumber\\
&=&-\frac{\kappa_5^2}{3}\int dy \textrm{e}^{-4A}\frac{(\partial_y A)^2}{\partial_y^2 A}<\infty.
\end{eqnarray}
To obtain the last equation, we used the Einstein equation \eqref{eqy1}. For a background geometry that is asymptotically AdS$_5$, such as our solution in eq.~\eqref{warpfactorSolution}, the integrand of the above integration is obviously divergent at $y=\pm\infty$. As a result, the above integration is divergent, and therefore, there is no localizable scalar zero mode in our model.

A localizable scalar zero mode corresponds to a new long range force gauge boson, which transmits a new force we have never seen before, and  therefore, is phenomenology unacceptable. It was pointed in~\cite{George2011} that a normalizable zero mode survives in five-dimensional models with two scalars constructed using a superpotential, even in the presence of warped gravity.

\subsection{Massive resonance modes}
In addition to the zero modes, we have a continuum of massive modes in both tensor and scalar sectors. These massive modes would modify gravity at small scale~\cite{CsakiErlichHollowoodShirman2000,Giovannini2003}. In this subsection, we argue that the structure of massive graviton modes can be largely different when $\beta$-term is introduced. For example, gravity resonances would emerge in both tensor and scalar sectors, if one switch on {a large $\beta$-term}. Let us start with the case without $\beta$-term.
\subsubsection{Without $\beta$-term}
In the standard model of thick brane ($\beta=0$), there is no gravity resonance. To see the absent of resonance in tensor section, it is more convenient to study the following equation:
\begin{eqnarray}
\label{SUSYtensorFactor}
\mathcal{J}^\dagger\mathcal{J}\tilde{\chi}=\tilde{m}^2\tilde{\chi}.
\end{eqnarray}
This is the dynamical equation of $\chi$'s superpartner, viz, $\tilde{\chi}$.
According to supersymmetric quantum mechanics, superpartners share the same spectrum except the ground state. Thus, if $\tilde{\chi}$ has massive resonant peaks, so does $\chi$. Expanding eq.~\eqref{SUSYtensorFactor}, we obtain another Schr\"odinger-like equation with the following potential:
\begin{eqnarray}
\tilde{U}_T(r) = \frac{9}{4}A'^2 - \frac{3}{2}A''.
\end{eqnarray}

\begin{figure}
\begin{center}
\includegraphics[width=0.5\textwidth]{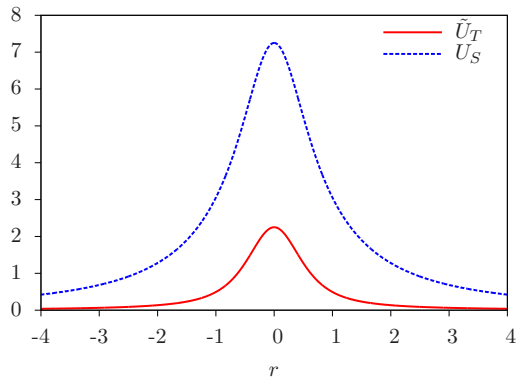}
\end{center}
\caption{$\tilde{U}_T$ and $U_S$ in standard model ($\beta=0$). We have take $k=1$. This figure shows that in the standard model of thick brane, there is no gravity resonances in both tensor and scalar sectors.} \label{figureSM}
\end{figure}

The numerical plot of $\tilde{U}_T(r)$ (figure~\ref{figureSM}) does not show any attractive well, so, it is impossible for $\tilde{\chi}$ to have resonant modes, so dose $\chi$.

Similarly, the plot of the scalar Schr\"odinger potential $U_S=\zeta\left(\zeta^{-1}\right)''$ also indicates the absent of gravity resonance in the canonical model (see also figure~\ref{figureSM}).

\subsubsection{$\beta$-term and gravity resonances}

\begin{figure*}
\begin{center}
\includegraphics[width=0.8\textwidth]{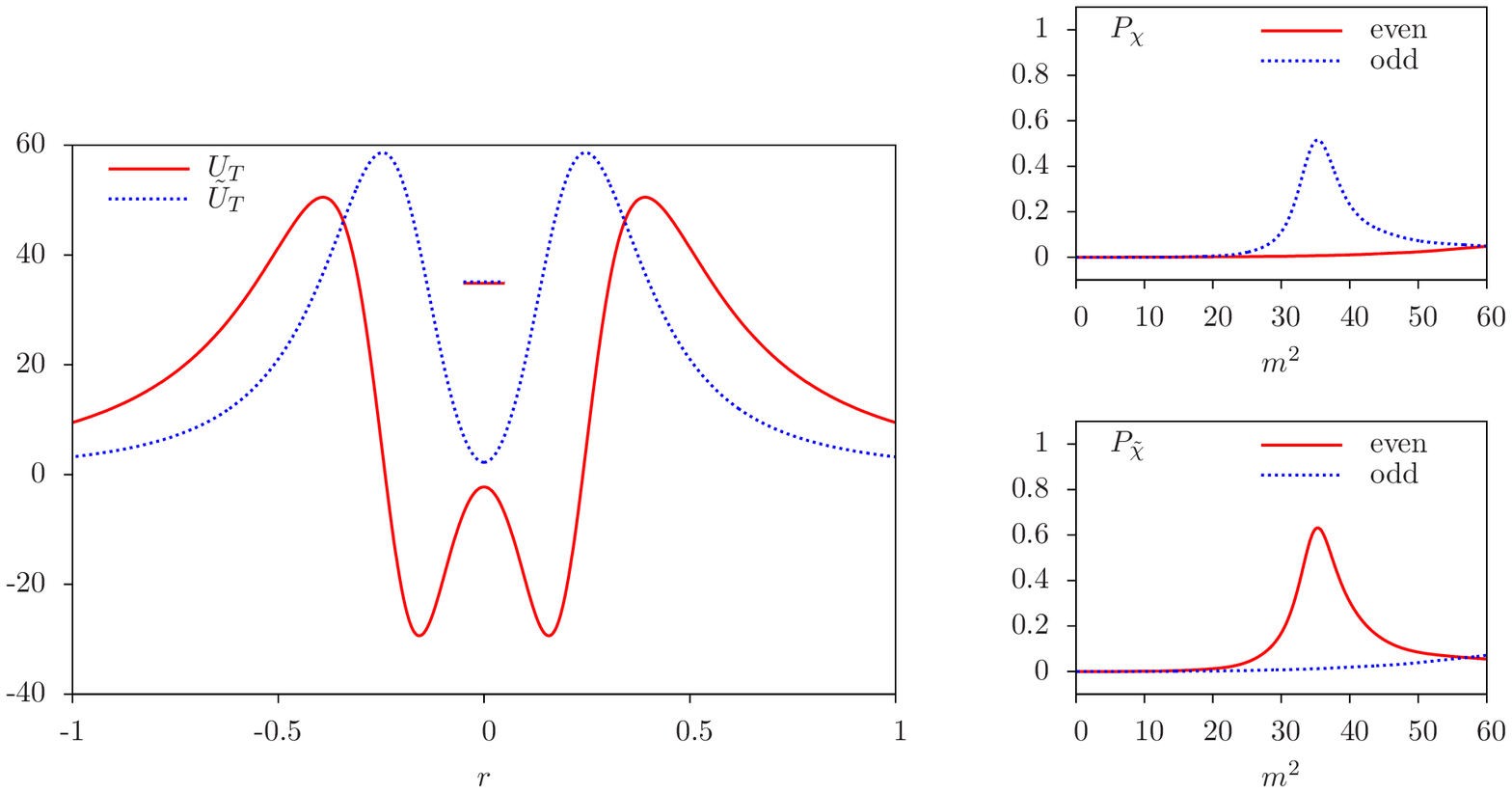}
\end{center}
\caption{The left panel is $U_T$ and $\tilde{U}_T$ with $n=1$, the right panel is the relative probability for $\chi_n$ (the upper one) and $\tilde{\chi}_n$ (the lower one). The parameters are $k=1$, $\phi_0=5$, and $\beta=35$.}
\label{figureTensorPert1}
\end{figure*}

\begin{figure*}
\begin{center}
\includegraphics[width=0.9\textwidth]{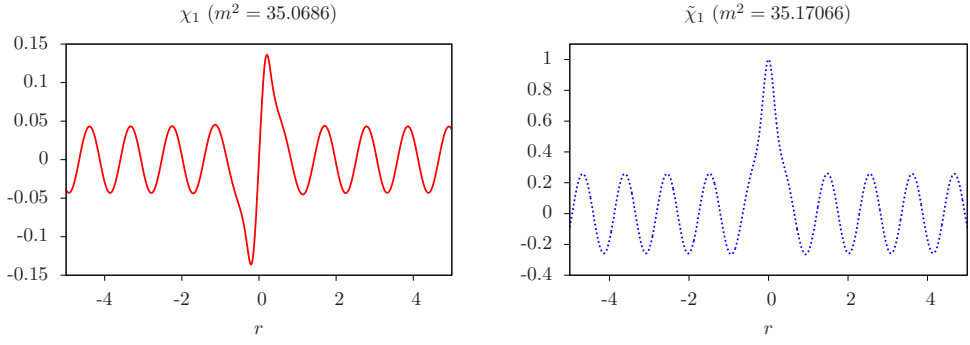}
\end{center}
\caption{Tensor wave function for $n=1$. The parameters are $k=1$, $\phi_0=5$, and $\beta=35$.} \label{figureTensorWaveFunction1}
\end{figure*}

\begin{figure*}
\begin{center}
\includegraphics[width=0.8\textwidth]{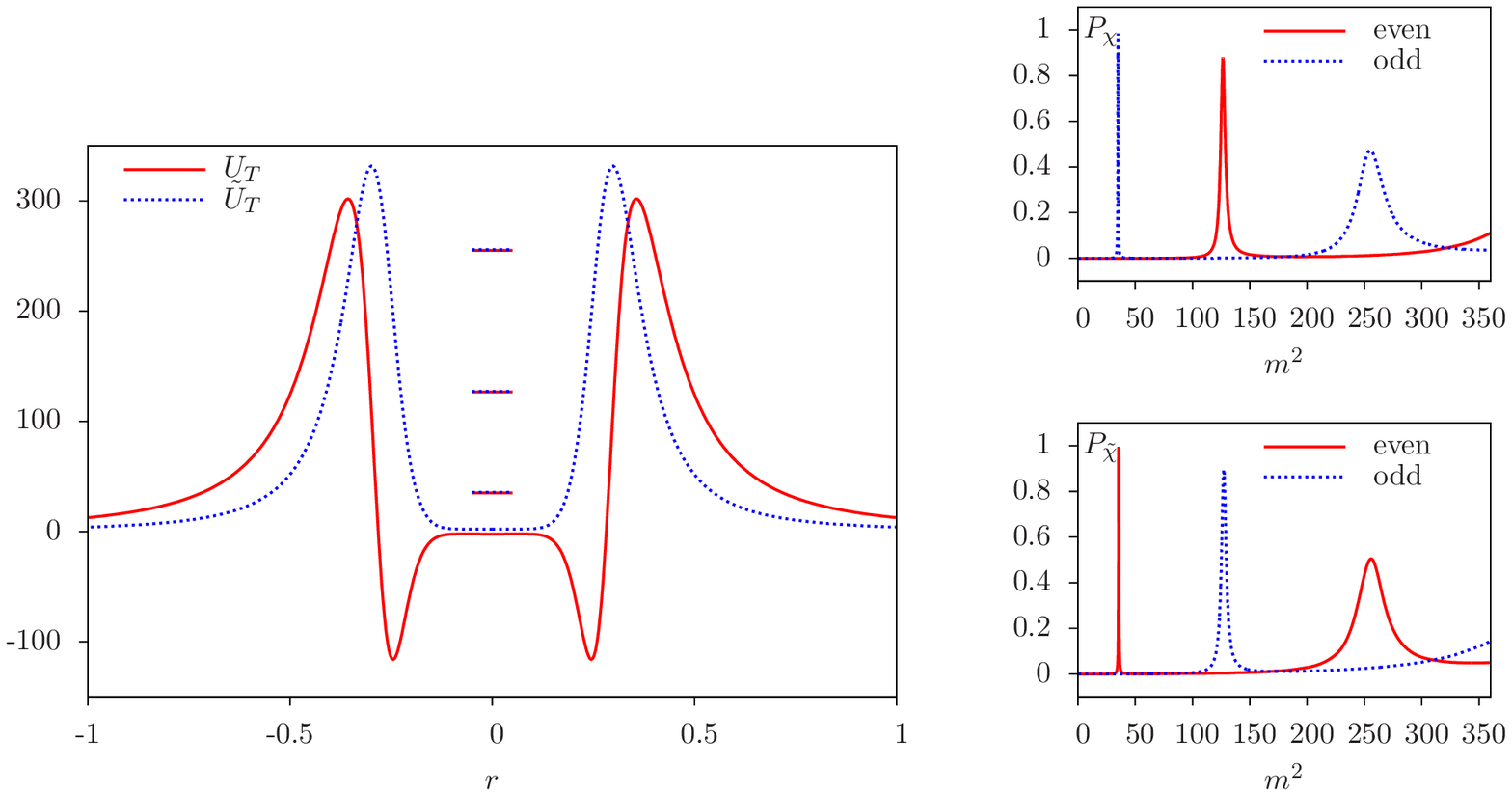}
\end{center}
\caption{The left panel is $U_T$ and $\tilde{U}_T$ with $n=4$, the right panel is the relative probability for $\chi_j$ (the upper one) and $\tilde{\chi}_j$ (the lower one). The parameters are $k=1$, $\phi_0=5$, and $\beta=35$.} \label{figureTensorBeta4}
\end{figure*}

\begin{figure*}
\begin{center}
\includegraphics[width=1\textwidth]{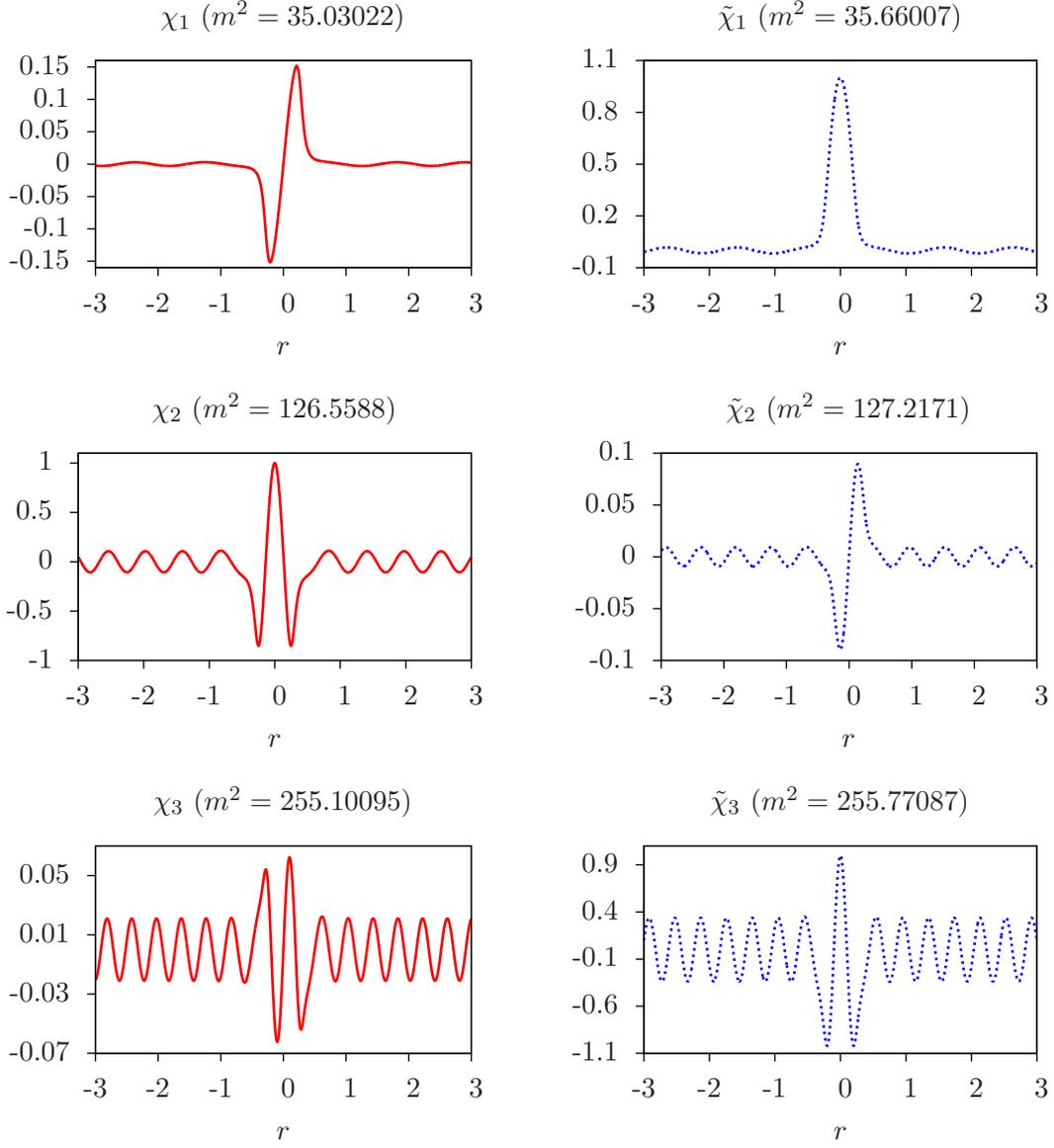}
\end{center}
\caption{Wave functions of the tensor resonant modes and their corresponding superpartners in the case with $n=4$, $k=1$, $\phi_0=5$, and $\beta=35$.} \label{figureTensorWaveFunction4}
\end{figure*}

\begin{table}[h]
\begin{center}
\rowcolors{2}{blue!8}{white}
\begin{tabular}{c|ccccc}
\rowcolor{gray!8}
\hline
$n$   &$\chi_i/\tilde{\chi}_i$& $m^2$  & $m$  & $\Gamma$   & $\tau$  \\ \hline
1     &$\chi_1$  & 35.06865  & 5.92188 & 0.563159  & 1.7757 \\
 & $\tilde{\chi}_1$& 35.17066& 5.93049 & 0.582229 &1.71754  \\ \hline\hline
      & $\chi_1$   & 35.03022 & 5.91863  & 0.0350 & 28.53288 \\
& $\tilde{\chi}_1$ & 35.66007 & 5.97161  & 0.0356 & 28.09282 \\
4  & $\chi_2$  & 126.5588 & 11.24983  & 0.2105435  & 4.74961  \\
 &$\tilde{\chi}_2$ & 127.2171 &11.27906& 0.178881& 5.59031 \\
  & $\chi_3$  & 255.10095   & 15.97188  & 0.8931337  & 1.11965  \\
  & $\tilde{\chi}_3$ & 255.77087 & 15.99284 & 0.911223 & 1.09743 \\ \hline
\end{tabular}
\end{center}
\caption{Resonant modes for tensor perturbation and its superpartner.
The parameters are $k=1$, $\phi_0=5$, and $\beta=35$.}
\label{tabTensorResonant}
\end{table}

\begin{figure*}
\begin{center}
\includegraphics[width=1\textwidth]{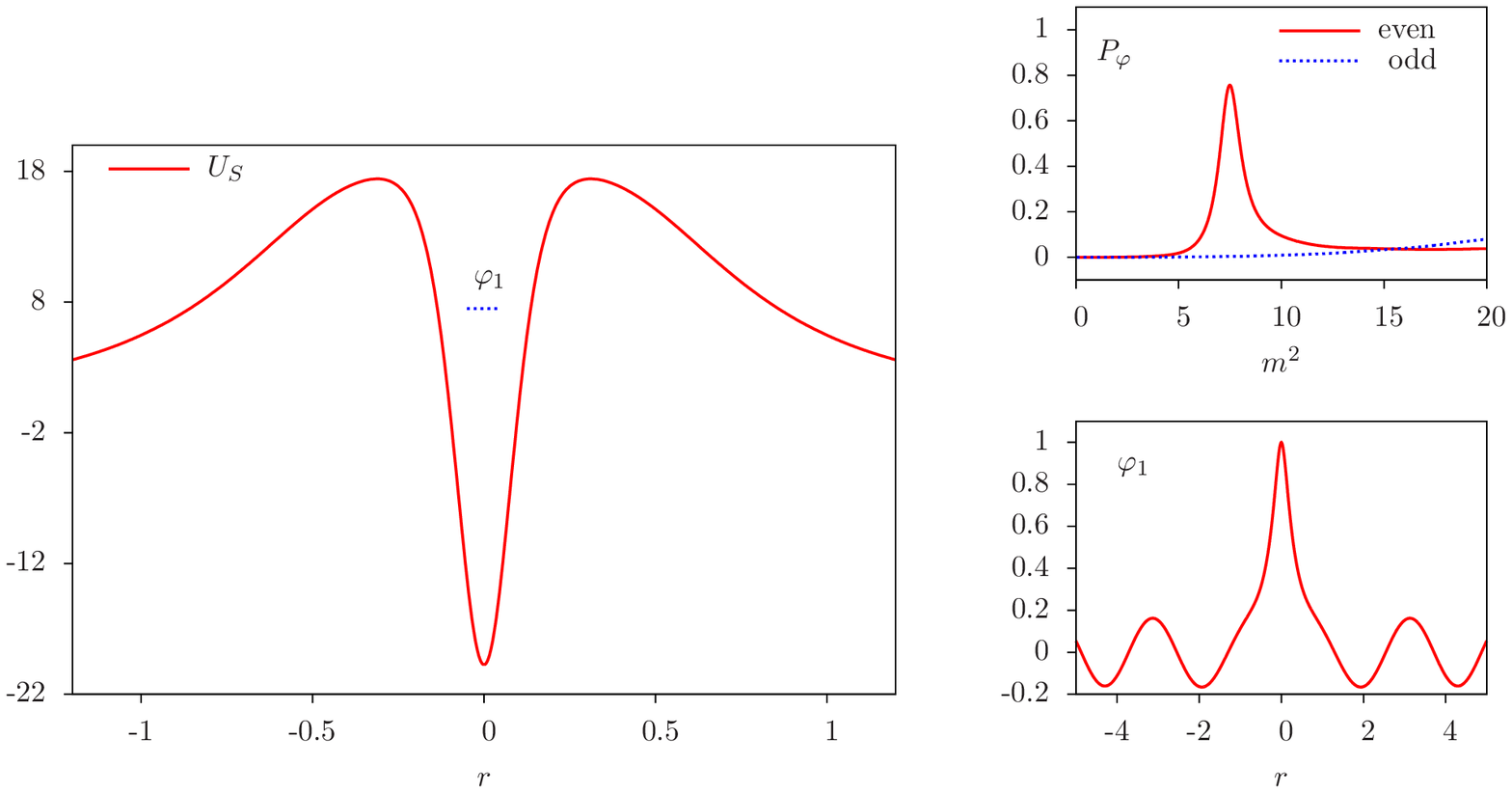}
\end{center}
\caption{Plots of the Schr\"odinger potential for scalar perturbation $U_s(r)$, the relative probability $P_\varphi$, and the wave function of the resonant mode. The parameters are $n=1$, $k=1$, $\phi_0=3$, and $\beta=3$.}
\label{figureScalarPotential1}
\end{figure*}

\begin{figure*}
\begin{center}
\includegraphics[width=1\textwidth]{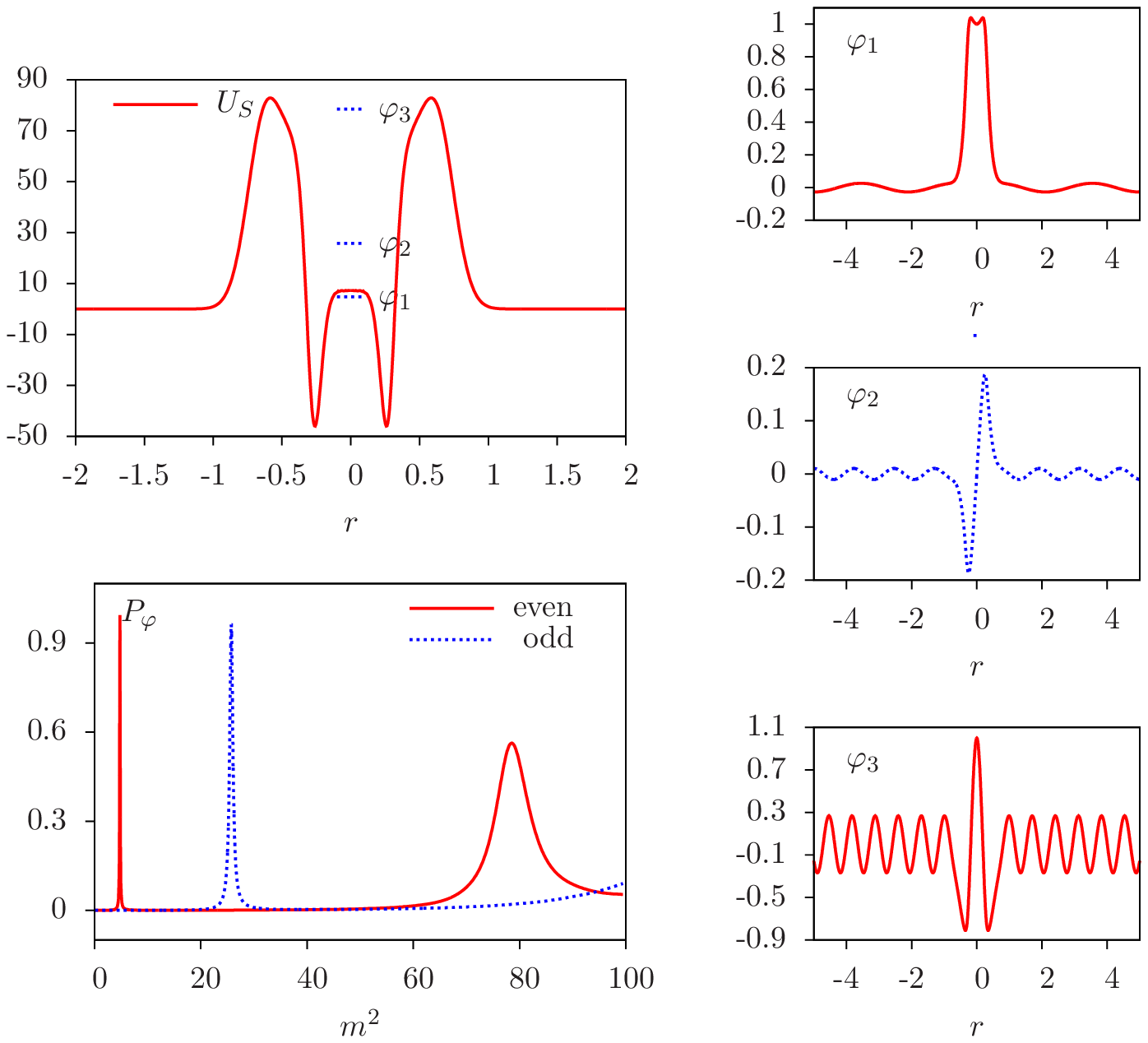}
\end{center}
\caption{Plots of the Schr\"odinger potential for scalar perturbation $U_S(r)$, the relative probability $P_\varphi$, and the wave functions of the resonant modes. The parameters are $n=4$, $k=1$, $\phi_0=3$, and $\beta=3$.} \label{figureScalarPotential4}
\end{figure*}

\begin{table}[h]
\begin{center}
\begin{tabular}{c|ccccc}
\rowcolor{gray!10}
\hline
$n$   &$\varphi_i$& $m^2$  & $m$  & $\Gamma$   & $\tau$  \\ \hline
1     &$\varphi_1$  & 7.49974  & 2.73857 & 0.2368  & 4.2238 \\ \hline\hline
      & $\varphi_1$   & 4.78727 & 2.18798  & 0.02734446 & 36.57048 \\
4  & $\varphi_2$  & 25.74273 & 5.07373  & 0.071041632  & 14.07625  \\
  & $\varphi_3$  & 78.501   & 8.86  & 0.4364411  & 2.29126  \\ \hline
\end{tabular}
\end{center}
\caption{Resonant modes for scalar perturbation.
The parameters are $k=1$, $\phi_0=3$, and $\beta=3$.}
\label{tabScalarResonant}
\end{table}

Now, let us come back to the noncanonical model. In figure~\ref{figureTensorPert1}, we plotted $U_T$ and $\tilde{U}_T$ {in the case of} $n=1$. Taking $\phi_0=5$ and $\beta=35$, we obtain a volcanolike potential $\tilde{U}_T$. Such an $\tilde{U}_T$ supports a continuum mass-square spectrum $\tilde{m}^2\in(0,+\infty)$. However, due to the barriers near $y=0$, some modes in the continuum {are} special, namely, the resonance modes: their wave functions are dominantly distributed around $y=0$. These modes are quisilocalized, after a finite time, they tunnel into the extra dimension.

In order to identify the resonance modes, we note that $\tilde{\chi}(r)$ can be regarded as the wave function in quantum mechanics. Then,
$|\tilde{\chi}_{m}(r)|^2 dr$ (after normalizing $\tilde{\chi}_m({r})$) can  be interpreted as the
probability for finding a KK mode with mass $m$ in an infinitesimal space range $(r,r+dr)$. Although none of
the massive modes is normalizable, one can still define
the function~\cite{LiuYangZhaoFuDuan2009}
\begin{eqnarray}
P(m)=\frac{\int^{r_b}_{-r_b}|\tilde{\chi}_m(r)|^2dr}
{\int^{10r_b}_{-10r_b}|\tilde{\chi}_m({r})|^2dr}
\end{eqnarray}
as the relative probability for finding a massive KK mode with mass $m$ in a
narrow range $r\in[-r_b, r_b]$ as compared to a wider interval
$r\in[-10r_b ,10r_b]$. Usually, we take
$2r_b$ as the width of the thick brane. Then $P(m)$ tends to $0.1$ when $m^2\gg U_S^{\rm{max}}$.
This is because $\varphi_m({r})$ can be approximately
identified as a plane wave for very large $m^2$.

In order to calculate $P(m)$ numerically, we take~\cite{LiuYangZhaoFuDuan2009}:
\begin{eqnarray}
{\varphi_m^{\rm{even}}(0)=1,\quad \partial_r\varphi_m^{\rm{even}}(0)=0,}
\end{eqnarray}
and
\begin{eqnarray}
{\varphi_m^{\rm{odd}}(0)=0,\quad \partial_r\varphi_m^{\rm{odd}}(0)=1,}
\end{eqnarray}
as the initial conditions for the even
and odd parity modes of $\varphi_m(z)$, respectively.

The numerical calculation for $n=1$, $k=1$, $\phi_0=5$, and $\beta=35$ shows that there is a resonant peak $\tilde{\chi}_1$ at $m^2=35.17066$. The appearance of the resonant mode in spectrum of $\tilde{\chi}$ reminds us that there should also be a corresponding mode (with the same mass) in the spectrum of $\chi$. The numerical {result} confirms our hypothesis: the corresponding resonant mode $\chi_1$ appears at $m^2=35.0686$ (see figures~\ref{figureTensorPert1}). The superpartners of {the graviton resonant modes} have nearly the same {masses}.

When $\phi_0>2$, the number of tensor resonances increases {with $n$}. For example, when we take $n=4$, $k=1$, $\phi_0=5$, and $\beta=35$, we obtain three resonances (see figure~\ref{figureTensorBeta4}). For each resonant peak, we define $\Gamma=\delta m$ as the mass width at half maximum of the corresponding peak. Then, the lifetime $\tau$ for a resonance is defined as $\tau=\Gamma^{-1}$. After a time $\tau$, these metastable gravitons decay into the extra dimension. The data for the tensor resonances is listed in table~\ref{tabTensorResonant}.

The wave function of $\chi$ and $\tilde{\chi}$ are plotted in figure~\ref{figureTensorWaveFunction1} for $n=1$, and figure~\ref{figureTensorWaveFunction4} for $n=4$. Comparing the behaviour of the wave functions of $\tilde{\chi}_n$ and $\chi_n$, we find that the $n$th resonant mode of $\tilde{U}_T$ behaves similarly as the $(n+1)$th resonant mode of $U_T$. This feature is one of the predictions of supersymmetric quantum mechanics, namely, superpartner potentials $\tilde{U}_T$ and $U_T$ have the same spectrum except for the zero energy ground state~\cite{CooperKhareSukhatme1995}.

Similarly, from eq.~\eqref{Psi}, we can analyze the resonant modes in the scalar sector. The numerical results are depicted in {figures} \ref{figureScalarPotential1} and \ref{figureScalarPotential4} and {table} \ref{tabScalarResonant}. As the tensor sector, the number of scalar resonances increases {with $n$}.

\section{Back to the standard frame}
\label{sec5}
In this section, we argue that the aforementioned noncanonical brane solution and its linear spectrum can also be obtained in a model with canonical dynamics. Obviously one can rewrite a noncanonical scalar lagrangian density
\begin{eqnarray}
\mathcal{L}(\phi)=G(\phi)X-U(\phi),
\end{eqnarray}
into the canonical form
\begin{eqnarray}
\mathcal{L}(\tilde{\phi})=-\frac12(\partial_y\tilde{\phi})^2-U(\tilde{\phi}),
\end{eqnarray}
by redefining the scalar field
\begin{eqnarray}
\label{phitransf}
\tilde{\phi}&=&\int d \phi\sqrt{G(\phi)}.
\end{eqnarray}
In our model, $G(\phi)=1+\beta\phi^{2n}$, so
\begin{eqnarray}
\tilde{\phi}&=& \frac{n\phi }{1+n} F_2^1\left[\frac{1}{2},\frac{1}{2 n},1+\frac{1}{2 n},-\beta  \phi ^{2 n}\right]\nonumber\\
&+&\frac{\phi \sqrt{1+\beta  \phi ^{2 n}}}{1+n}.
\end{eqnarray}
Obviously, when $\beta=0$, we have $\tilde{\phi}=\phi$.
\begin{figure*}
\begin{center}
\includegraphics[width=1\textwidth]{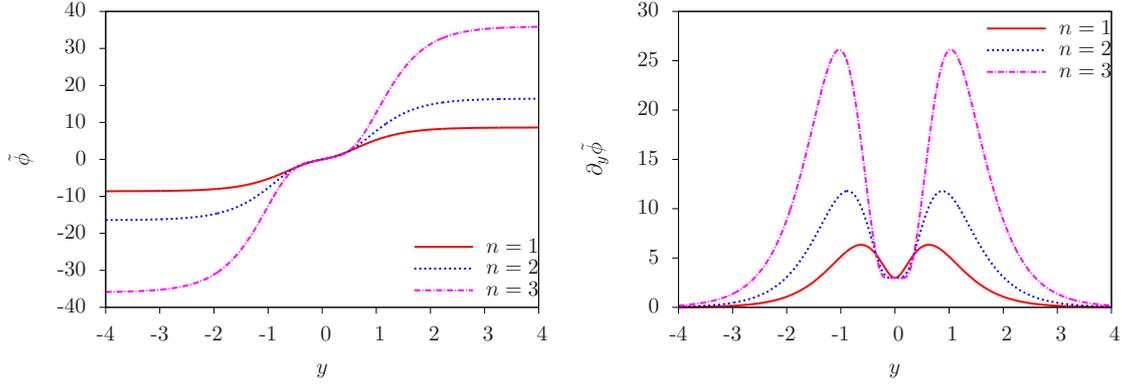}
\end{center}
\caption{The scalar field $\tilde{\phi}$ and its first-order derivative for $n=1,2,3$. The parameters are $k=1$, $\phi_0=3$, and $\beta=3$.} \label{figureTildePhiandPrime}
\end{figure*}
From figure~\ref{figureTildePhiandPrime} we see that when $\beta>0$, the configuration of scalar field $\tilde{\phi}$ undergoes a deformation as $n$ increases.

This feature of $\tilde{\phi}$ reminds us a similar solution of ref.~\cite{BazeiaFurtadoGomes2004}, in which the brane deformation is realized in the canonical frame by taking a special superpotential:
\begin{eqnarray}
\tilde{W}_p=\frac{{2p}}{{2p - 1}}{\tilde{\phi} ^{(2p - 1)/p}} - \frac{{2p}}{{2p + 1}}{\tilde{\phi} ^{(2p + 1)/p}}.
\end{eqnarray}
The parameter $p=1,3,5\cdots$ is an odd integer. As $p$ increases, the configuration of $\tilde{\phi}$ undergos a transition from single kink to double kink\footnote{Note that the $\tilde{\phi}$ in this paper is not a {standard} double kink solution, because $\partial_y\tilde{\phi}$ is not vanished at $y=0$ (see the right panel of figure~\ref{figureTildePhiandPrime}).}. However, the authors of ref.~\cite{BazeiaFurtadoGomes2004} did not find any sign of gravity resonance for $p=1,3,5$. As $p$ increases, the ability for trapping graviton deceases.

In order to derive the corresponding canonical superpotential $\tilde{W}(\tilde{\phi})$ of our solution, let us note that the first-order equations in the canonical frame are:
\begin{eqnarray}
\partial_y A&=&-\frac{\kappa_5^2}{3}\tilde{W},\\
\label{phiWSM}
\partial_y \tilde{\phi}&=&\frac{d\tilde{W}}{d\tilde{\phi}},
\end{eqnarray}
and that $\phi$ and $\tilde{\phi}$ are related by the following equation:
\begin{eqnarray}
\frac{d\tilde{\phi}}{d\phi}=\sqrt{G(\phi)}.
\end{eqnarray}
Then from eq.~\eqref{phiWSM} we obtain
\begin{eqnarray}
\tilde{W}&=&\int_0^{\tilde{\phi}}G(\phi) W_\phi d\phi.
\end{eqnarray}
{So,} given an arbitrary superpotential $W(\phi)$ in the noncanonical frame, we can always obtain the corresponding superpotential $\tilde{W}(\tilde{\phi})$ in the canonical frame. Especially, for our solution, we have
\begin{eqnarray}
\tilde{W}&=&k \phi _0^2\left[\frac{\tilde{\phi} }{\phi_0}-\frac{1}{3}\left(\frac{\tilde{\phi} }{\phi _0}\right)^3\right]
+k \phi _0^2\left[\frac{\beta  \tilde{\phi} ^{2n}}{1+2 n}\frac{\tilde{\phi} }{\phi _0}-\frac{\beta  \tilde{\phi} ^{2n}}{(3+2 n)}\left(\frac{\tilde{\phi} }{\phi _0}\right)^3\right].
\end{eqnarray}
The scalar potential can be obtained from the following equation
\begin{eqnarray}
U(\tilde\phi) = \frac{1}{2}\tilde{W}_\phi ^2  - \frac{2}{3}\kappa _5^2{\tilde{W}^2}.
\end{eqnarray}

Now, let us comment that the canonical and the noncanonical solutions generated by $\tilde{W}(\tilde{\phi})$ and $W(\phi)$, correspondingly, possess the same linear spectrum. On one hand, the scalar perturbation equation in the canonical frame can be obtained from the noncanonical equation \eqref{schro} by simple taking $\mathcal{L}_X=1$ and replace $\phi\to\tilde{\phi}$, so that
\begin{eqnarray}
\label{49}
\Psi =\textrm{e}^{-3A/2}\tilde{\phi}'\hat{\Psi},
\end{eqnarray}
and
\begin{eqnarray}
\label{50}
\zeta= \textrm{e}^{3A/2}\frac{\tilde{\phi}'}{A' }.
\end{eqnarray}
On the other hand, from eq.~\eqref{phitransf}, we have $\tilde{\phi}'=\sqrt{G}\phi '=\sqrt{\mathcal{L}_X}\phi '$. So, eqs.~\eqref{49} and \eqref{50} are nothing but eqs.~\eqref{30} and \eqref{31}. The tensor equation is independent of the scalar sector, so also remains unchanged. Therefore, all the graviton resonances we obtained in the noncanonical model can also be reproduced in the corresponding canonical model.

\section{Trapping massless fermion in canonical or noncanonical frame}
\label{sec6}
In a realistic brane model, matter fields must be trapped on the brane. In the standard model of thick brane, a massless left-handed fermion (the zero mode) can be localized on the brane, if one {assumes that} the bulk fermion $\Theta(x^\mu,r)$ is coupled with the background scalar. The simplest coupling is the Yukawa coupling $\eta\tilde{\phi}\bar{\Theta}\Theta$, where $\eta>0$ is the coupling constant.
Now, suppose that the thick brane is generated by the noncanonical scalar field $\phi$, then we would obtian a different localization condition for the fermion zero mode (because $\phi$ and $\tilde{\phi}$ are different if $\beta\neq0$). In this section, we study how the parameters $\beta$ and $n$ affect the localization of the fermion zero mode in both the noncanonical and the canonical frames.

Let us first consider the noncanonical model, and assume the action of $\Theta$ to be
\begin{eqnarray}
\label{Lfermion}
S_{1/2}=\int d^5x\sqrt{-g}\bar{\Theta}(\Gamma^M D_M-\eta \phi)\Theta.
\end{eqnarray}
Here, $\Gamma^M=(\textrm{e}^{-A}\gamma^\mu,~\textrm{e}^{-A}\gamma^5)$ and $D_M=\partial_M+\omega_M$ are the $\Gamma$-matrixes and covariant derivative in the five-dimensional curved space-time, respectively, $\omega_M=(\frac12A'\gamma_\mu\gamma_5,~0)$ is the spin connection (see ref.~\cite{LiuZhangZhangDuan2008} for details). From the action \eqref{Lfermion}, we immediately obtain the equation for $\Theta$:
\begin{eqnarray}
\label{EOM}
\{\gamma^\mu\partial_\mu+\gamma^5(\partial_r+2A')-\eta \textrm{e}^A\phi\}\Theta=0.
\end{eqnarray}
To continue, introducing the KK decomposition
\begin{eqnarray}
\label{theta}
\Theta=\textrm{e}^{-2A}\sum_C\sum_j\psi_{C,j}(x^\mu)f_{C,j}(r),
\end{eqnarray}
where $j$ denotes different excitations of the modes, and $C\in \{+,-\}$ denotes the chirality: $\psi_{+,j}$ and $\psi_{-,j}$ represent the right- and left-chiral modes, respectively.

Assuming that $\psi_{C,j}(x^\mu)$ are the chiral fermion that we observed in our four-dimensional world, we get
\begin{eqnarray}
\gamma^\mu\partial_\mu\psi_{C,j}(x^\rho)&=&m_j\psi_{-C,j}(x^\rho),\nonumber\\
\psi_{C,j}&=&C\gamma^5\psi_{C,j}.
\end{eqnarray}

Inserting eq.~\eqref{theta} into the equation of motion \eqref{EOM}, we obtain a Schr\"odinger-like equation for $f_{C,j}(r)$:
\begin{eqnarray}\label{Vf}
(-\partial_r^2+V_C(r))f_{C,j}=m_j^2f_{C,j},
\end{eqnarray}
{where} the potential is
\begin{eqnarray}
V_C&=&(\eta \textrm{e}^A \phi)^2+C\partial_r(\eta \textrm{e}^A \phi).
\end{eqnarray}
Defining $\mathcal{F}\equiv \partial_r+C\eta \textrm{e}^A \phi$, we can rewrite eq.~\eqref{Vf} as follows:
\begin{eqnarray}
\label{decompFermi}
\mathcal{F}\mathcal{F}^\dagger f_{C,j}=m^2_j f_{C,j}.
\end{eqnarray}
As we have stated in previous sections, the eigenvalues of such an equation are semi-positive definite, namely, $m^2_j\geq0$. Now, let us study the localization of the zero mode $f_{C,0}$, which satisfies a simpler equation:
\begin{eqnarray}
(-\partial_r+C\eta \textrm{e}^A \phi) f_{C,j}=0.
\end{eqnarray}
After an integration, one immediately {obtains }
\begin{eqnarray}
f_{C,0}(r)\propto \exp(C\eta\int_0^r d\bar{r}\textrm{e}^{A(\bar{r})}\phi(\bar{r})).
\end{eqnarray}
To trap the zero mode on the brane, we demand the integration $\int dr (f_{C,0})^2$ to be finite, or, when written in $y$-coordinate~\cite{LiuZhangZhangDuan2008}:
\begin{eqnarray}
\label{integral}
\mathcal{I}=\int dy \exp\left(-A(y)+2C\eta\int ^y_0 d\bar{y}\phi(\bar{y})\right)<\infty.
\end{eqnarray}
According to eqs.~\eqref{Aasyp} and \eqref{phi}, the asymptotical behaviour of the integrand is
\begin{eqnarray}
\left(k+ \beta c_n k+ 2 C\eta\phi_0\right)|y|,\quad \textrm{for} \quad |y|\to +\infty.
\end{eqnarray}
Obviously, the integral $\mathcal{I}$ converges only when
\begin{eqnarray}
\label{locali}
k+ \beta c_n k+ 2 C\eta\phi_0<0.
\end{eqnarray}
As the stability condition eq.~\eqref{stab} is considered, one can conclude that the left-chiral zero mode is localizable, provided
\begin{eqnarray}
\label{localiInequa}
\eta\phi_0>\frac k2(1+\beta c_n).
\end{eqnarray}
To estimate the value of $\eta$, let us take the parameters to be $n=4$, $k=1$, $\phi_0=5$ and $\beta=35$. With this set of parameters, we can get $3$ resonances in the tensor sector and at least $3$ in the scalar sector. Then, the localization condition for fermion zero mode is $\eta>41430$.

Now, let us turn to another possibility, namely, $\Theta$ is coupled with the canonical field $\tilde{\phi}$. The localization condition for this case can be obtained by simply replace $\phi_0$ in the left hand side of {eq.} \eqref{localiInequa} by
\begin{eqnarray}
\tilde{\phi}_0&=& \phi_0d_n,
\end{eqnarray}
where
\begin{eqnarray}
d_n&=&\frac{n}{1+n} F_2^1\left[\frac{1}{2},\frac{1}{2 n},1+\frac{1}{2 n},-\beta  \phi_0 ^{2 n}\right]\nonumber\\
&+&\frac{\phi_0 \sqrt{1+\beta  \phi_0 ^{2 n}}}{1+n}.
\end{eqnarray}
For parameters $n=4$, $k=1$, $\phi_0=5$ and $\beta=35$, the localization condition {reads as} $\eta>56$. To conclude, in order to have a same number of gravity resonance modes, and at the same time, to localize the fermion zero mode, the Yukawa coupling between $\Theta$ and $\phi$ might be hundreds times larger than the one between $\Theta$ and $\tilde{\phi}$.

\section{Conclusion}
In this paper, we successfully {constructed} a domain wall brane model with inner brane structure and graviton resonances by using a single scalar field $\phi$ in five-dimensional general relativity. The scalar field can be either noncanonical or canonical. In noncanonical frame, the inner brane structure emerges as one switch on the scalar-kinetic coupling. The number of the resonant modes depends on the vacuum expectation value of $\phi$ and the form of scalar-kinetic coupling. While in the canonical frame, the emergence of brane structure is caused by the deformation of $\tilde\phi$, which can be obtained from $\phi$ by doing an integration. With this relation, the canonical model and the noncanonical one share the same linear structure. So the gravity resonances {we obtained} in the noncanonical frame can also be obtained in the standard model. However, due to the inequivalence between the corresponding background scalar solutions, the localization condition for the left-chiral fermion zero mode can be largely different in different frames. Our estimate {showed} that the magnitude of the Yukawa coupling in the noncanonical frame might be hundreds times larger than the one in the canonical frame, if one demands the localization of the left-chiral fermion zero mode as well as the appearance of a few gravity resonance modes.

This work can be regarded as the first step for a further study on the phenomenological indications of the graviton resonances, which will be addressed in our future works.

\section*{Acknowledgments}
We thank Oriol Puj\`olas for reminding us the relation between the noncanonical and the canonical model, and Matteo Baggioli for many interesting discussions. Y. Zhong and Y.-X. Liu was supported by the Program for New Century Excellent Talents in University, the National Natural Science Foundation of China (Grants No. 11075065 and No. 11375075), and the Fundamental Research Funds for the Central Universities (Grant No. lzujbky-2013-18). Y. Zhong was also supported by the scholarship granted by the Chinese Scholarship Council (CSC). Z.-H. Zhao was supported by the National Natural Science Foundation of China (Grant No. 11305095) and the Natural Science Foundation of Shandong Province, China (Grant No. 2013ZRB01890), and the Scientific Research Foundation of Shandong University of Science and Technology for Recruited Talents (Grant No. 2013RCJJ026).

\providecommand{\href}[2]{#2}\begingroup\raggedright\endgroup
\end{document}